\documentclass[twocolumn,showpacs,preprintnumbers,amsmath,amssymb,prb]{revtex4} \newcommand\PrePrint{0}
\usepackage[dvips]{color,graphics,epsfig,rotating}
\usepackage{graphicx}
\graphicspath{{./}{figs/}}
\usepackage{dcolumn}
\usepackage{bm}
\usepackage{color}

\makeatletter
\ifthenelse{\equal{\PrePrint}{1}}
{
  \def\@dotsep{4.5}
}
{
}
\makeatother

\newcommand\eq[1]                              
{
\begin{align*}
#1
\end{align*}
}

\newcommand\eql[2] 
{
\begin{equation}\label{#1}
\begin{split}
#2
\end{split}
\end{equation}
}

\newcommand\eqsl[1]                            
{
\begin{align}
#1
\end{align}
}

\newcommand\Eq[1]      {Eq.~\eqref{#1}}

\newcommand\Ref[1]     {Ref.~\onlinecite{#1}}

\newcommand\eg[1]      {{\em e.g.}, #1}

\newcommand\ME[3]      {\langle{{#1}}|{{#2}}|{{#3}}\rangle} 
\newcommand\ket[1]     {|{{#1}}\rangle}

\newcommand\braket[2]  {\langle{{#1}}|{{#2}}\rangle}

\newcommand\PsiGS      {\Psi_0}

\newcommand\Hop        {{\hat{H}}}

\newcommand\GAMESS     {{\footnotesize{GAMESS}}}
\newcommand\NWCHEM     {{\footnotesize{NWCHEM}}}

\newcommand\Eh         {\ensuremath{E_\textrm{h}}}
\newcommand\mEh        {\textrm{m}\ensuremath{E_\textrm{h}}}

\newcommand\Cz         {C$_{\textrm{2}}$}

\definecolor{Green}{rgb}{0.2,0.96,0.2}
\definecolor{Remarks}{rgb}{1,0.3,0.3}
\definecolor{Extra}{rgb}{0.2,0.2,1}
\definecolor{Blue}{rgb}{0.2,0.3,1}
\definecolor{Black}{rgb}{0,0,0}

\newcommand\COMMENTED[1] {}

\newcommand\FIGDIR[1]   {figs/}

\begin{document}

\title{
Excited state calculations using phaseless auxiliary-field quantum Monte Carlo:
potential energy curves
of low lying C$_2$ singlet states
}

\author{Wirawan Purwanto}
\author{Shiwei Zhang}
\author{Henry Krakauer}
\affiliation{Department of Physics, College of William and Mary,
Williamsburg, Virginia 23187-8795, USA}

\date{\today}

\begin{abstract}

We show that the recently developed phaseless auxiliary-field quantum Monte
Carlo (AFQMC) method can be used
to study excited states, providing 
an alternative to
standard quantum chemistry methods.
The phaseless AFQMC approach, whose computational cost
scales as $M^3$-$M^4$ with system size $M$,  has
been shown to be among the most accurate many-body methods in ground state calculations.
For excited states,
prevention of collapse into the ground state and
control of the Fermion sign/phase problem are accomplished by the approximate phaseless
constraint with a trial
wave function. 
Using the challenging C$_2$ molecule as a test case,
we
calculate the potential energy
curves of the ground and
two low-lying singlet excited states.
The trial wave function is obtained by truncating complete active space
wave functions, with no further optimization. 
The phaseless AFQMC results using a small basis set are in
good agreement with exact full configuration interaction calculations, while those
using large basis sets are in good agreement with experimental spectroscopic
constants. 

\end{abstract}

\COMMENTED{
\begin{verbatim}
---------------------------------------------------------------------------
GAFQMC PAPER 2: C2 MOLECULE/EXCITED STATES
CVS $Id: NEW $
---------------------------------------------------------------------------



\end{verbatim}
}


\maketitle

\section{Introduction}

The ability to calculate electronic excited states of molecules and extended systems
is necessary to predict key phenomena and properties of technologically important systems.
Compared to ground states, however, the accurate calculation of excitation energies is significantly more difficult.
For molecules,
a variety of many-body electronic structure quantum chemistry
approaches have been developed, 
typically using a one-particle basis
to represent the many-body wave function. For small molecules with
modest basis sets, the full configuration-interaction (FCI) method is
exact, but FCI is not practical for realistic calculations, since the
computational cost scales exponentially as the system
size is increased. For larger 
systems, 
approximate coupled cluster (CC)
methods \cite{bartlett-RMP2007} are the standard, but these methods also have rather steep
computational scaling with system size [\eg $O(M^7)$ for
CCSD(T), CC with single and double excitations and perturbational
triplets, where $M$ is the number of basis functions].
For extended systems, less accurate approximations based on density functional theory (DFT) and
time-dependent DFT 
have been developed; GW and Bethe-Salpeter type
methods have also been shown to be promising. \cite{Excitations_GreensFnct_DFT}
Correlated quantum chemistry methods have also been embedded in DFT calculations to treat
extended systems. \cite{2002-embed-excit-Carter,2007-mb-pbc-Scheffler}
The most commonly applied quantum Monte Carlo (QMC) method in electronic
structure has been diffusion Monte Carlo (DMC),\cite{Foulkes2001} which
has also been used
to compute excited states.
\cite{1999-symmetry-DMC-excited,2001-dmc-mol-excit-Grossman}
Compared to ground states, however, the accuracy of the results may depend on
the symmetry \cite{1999-symmetry-DMC-excited}
and show greater sensitivity~\cite{2004-excit-DMC-filippi-2}
to the trial wave function
used in the fixed-node approximation
to control the sign problem and maintain orthogonality.

The recently developed phaseless auxiliary-field quantum Monte Carlo
(AFQMC) method
\cite{SZ-HK:2003,Al-Saidi_TiO-MnO:2006,Al-Saidi_GAFQMC:2006,QMC-PW-Cherry:2007}
is an orbital-based alternative many-body approach.  AFQMC can
be expressed with respect to any chosen single-particle basis
(\eg{gaussians, planewaves, Wannier, etc.}), and it
exhibits favorable $O(M^3$-$M^4)$ scaling.
For ground states,
the new AFQMC method has been applied to close to 100 systems, including
first- and second-row molecules, \cite{Al-Saidi_GAFQMC:2006,QMC-PW-Cherry:2007,AFQMC-CPC2005}
transition metal oxide molecules, \cite{Al-Saidi_TiO-MnO:2006}
simple solids, \cite{SZ-HK:2003,2008-FS-Hendra}
post-$d$ elements, \cite{Al-Saidi_Post-d:2006} van der Waals
systems, \cite{Al-Saidi_H-bonded:2007}
and in molecules in which bonds are being stretched or
broken. \cite{Al-Saidi_Bondbreak:2007,Purwanto2008} In these
calculations we have operated largely in an automated mode, inputting only the
DFT or Hartree-Fock (HF) solutions as trial wave functions. 
The method demonstrated excellent accuracy, consistently able to correct errors
in the mean-field trial wave function.
In molecules,
we have found that the accuracy of the phaseless AFQMC is comparable to CCSD(T)
near equilibrium geometry and better when bonds are stretched. 
AFQMC thus provides new
opportunities for the efficient and accurate many-body calculations of ground and excited states
in both molecular and extended systems.

The seemingly simple {\Cz} molecule presents a significant
challenge for many-body methods. \cite{Abrams2004,2007-WF-optimiz}
The {\Cz} molecule is difficult
because of the strongly multireference nature of the ground state wave
function [in which only $\sim 70$\% of the weight is the restricted
Hartree-Fock (RHF) determinant] and the presence of nearby low-lying states.
The shortcomings of standard quantum chemistry calculations for {\Cz}
were shown by recent benchmark FCI calculations~\cite{Abrams2004} of
the potential energy curves (PECs) of its
$^1\Sigma_g^+$ ground state and two low-lying
singlet excited states.
This benchmark shows that most correlated methods based on a
single-determinant reference state wave function $\ket{\Phi_r}$ exhibit
large nonparallelity errors (NPE---defined as the difference between the
maximum and minimum deviations from FCI along the PEC).
Spin-restricted CCSD(T) 
[referred to as RCCSD(T) hereafter]
was found to exhibit a large NPE of $98\,\mEh$
due to the poor behavior of RHF in the dissociation limit.
Spin-unrestricted UCCSD(T), 
which is usually less accurate near equilibrium, has an NPE of $34\,\mEh$.
The excited state PECs are not
accurately modeled by any of the commonly used single-reference
methods, nor by CI including full quadruple
substitutions. \cite{Abrams2004}
Similarly, 
a recent DMC study \cite{2007-WF-optimiz}
found that, even in its ground state at equilibrium
geometry, the total
energy of {\Cz} showed a large fixed-node error $\sim 40\,\mEh$, if
a single-determinant trial wave function is used.
%

As a new QMC method, the phaseless AFQMC provides an alternative route to
the sign problem from fixed-node DMC.
The random walks take place in a manifold of
Slater determinants, in which fermion antisymmetry is automatically maintained
in each walker.
Applications have indicated that often this
reduces the severity of the sign problem and, as a result, the phaseless
approximation has weaker reliance on the trial wave function.
It is interesting then to test the method for excited states, where
QMC calculations depend more on the trial wave function and our knowledge of
it is less.
The challenging
{\Cz} molecule, where FCI results are available for the modest-sized basis set
6-31G*,
provides an excellent test case.

We first describe the AFQMC methodology for ground and excited states.
We then make detailed comparions of our  {\Cz} calculated results with
the FCI
calculations of \Ref{Abrams2004}.
Finally, our calculated PECs and spectroscopic
constants with large realistic basis sets are presented and
compared with experimental results.

\section{Methodology}
\label{sec:method}

In this paper, we focus on the
lowest-lying {\Cz} singlet states: the $X\,^1\Sigma_g^+$ ground state
and the $B\,^1\Delta_g^+$ and $B'\,^1\Sigma_g^+$ excited states.
Since AFQMC uses a projection method to obtain the excited states,
collapse to the ground state must be prevented. The
$B\,^1\Delta_g^+$ state belongs to a different irreducible
representation of the symmetry group of the Hamiltonian than does the
$X\,^1\Sigma_g^+$ ground state, but the $B'\,^1\Sigma_g^+$ excited
state belongs to the same irreducible representation as the ground
state. Both of these cases are discussed below. We first briefly
review the phaseless AFQMC method and then discuss the calculation
of the excited states.

\subsection{Ground state}
\label{sec:GS}

Stochastic ground state quantum Monte Carlo (QMC) methods,
\cite{Ceperley1980,Reynolds1982,Foulkes2001,SZ-HK:2003} which are
exact in principle, use projection from any reference many-body wave
function $\ket{\Phi_r}$, which has non-zero overlap with the ground
state.  In practice, however, the Fermionic sign problem
\cite{Ceperley_sign,kalos91,Zhang1999_Nato,Foulkes2001,SZ-HK:2003}
must be controlled to eliminate exponential growth of the variance.  For example, in DMC,
a single- or multi-reference trial wave function 
is used to impose approximate
nodal boundary conditions of the many-body wave function in electronic
configuration space (a
Jastrow factor is also included to reduce the stochastic variance). By contrast, phaseless AFQMC
samples the many-body wave function with random walkers
$\{ \ket{\phi} \}$
in the space of Slater determinants,
which are expressed in terms of a
chosen single-particle basis. Here we use standard quantum chemistry gaussian basis
sets. \cite{EMSL_BasisSets2007}
Each $\ket{\phi}$ has the form of a HF or DFT wave function,
with the orbitals varying stochastically in the projection.
AFQMC 
controls the sign problem differently, using the complex overlap of the walker $\ket{\phi}$
with a trial/reference wave function $\ket{\Phi_r}$ which is a
determinant or a linear combination of determinants.

The ground state energy is obtained from the mixed
estimator
\begin{equation}
E_0 = \frac{\ME{\Phi_r}{\Hop}{\PsiGS}}
               {\braket{\Phi_r}{\PsiGS}}
    =  \lim_{\beta \to \infty}
          \frac{\ME{\Phi_r}{\Hop e^{-\beta\Hop}}{\Phi_r}}
               {\ME{\Phi_r}{e^{-\beta\Hop}}{\Phi_r}}
        \,,
\label{eq:Emixed}
\end{equation}
where $\Hop$ is the many-body Hamiltonian and $\ket{\PsiGS}$ is the ground state wave function, which
is given by imaginary time ($\beta$) projection from $\ket{\Phi_r}$.
Using the Hubbard-Stratonovich (HS) transformation, \cite{Hubbard,Stratonovich}
an importance sampling transformation \cite{SZ-HK:2003} then
expresses the mixed estimator as a
stochastic average over the walkers and their Monte Carlo weights $w_\phi$,
\begin{equation}
    E_0^{\textrm{MC}}
  = \frac{\sum_\phi w_\phi E_L[\phi]}{\sum_\phi w_\phi}
    \,,
\label{eq:mixed_w_EL}
\end{equation}
where the ``local energy'' $E_L$ is defined as
\begin{equation}
    E_L[\phi]
    \equiv
    \frac{\ME{\Phi_r}{\Hop}{\phi}}
         {\braket{\Phi_r}{\phi}}
    \,.
\label{eq:El}
\end{equation}

The energy computed from Eq.~(\ref{eq:mixed_w_EL}) is approximate. In addition
to a statistical error which can be accurately estimated 
and reduced with further sampling, there is a systematic error because of
the phaseless constraint with $|\Phi_r\rangle$. In other words, the
realization of
$e^{-\beta\Hop}$ in Eq.~(\ref{eq:Emixed}) is approximate,
\begin{equation}
e^{-\beta\Hop} \rightarrow \widetilde{e^{-\beta\Hop}}\,,
\label{eq:prop_appr}
\end{equation}
because of
the constraint on the random walk paths in the Slater determinant space (or
equivalently, in the corresponding auxiliary-field space).
This is the only approximation in the calculation. The computed ground
state energy is not an upper bound. \cite{SZ-HK:2003,sz-cpmc-1999}

In previous applications, phaseless AFQMC with a single
unrestricted Hartree-Fock (UHF) determinant
$\ket{\Phi^{\textrm{UHF}}_r}$ was found to often give better overall
and more uniform accuracy than CCSD(T) in mapping PECs.
\cite{Al-Saidi_GAFQMC:2006,Al-Saidi_Post-d:2006,Al-Saidi_H-bonded:2007}
In some cases, however, such as the BH and N$_2$ molecules, achieving
quantitative accuracy of a few $\mEh$ for the entire PEC required
multi-determinant $\ket{\Phi_r}$. \cite{Al-Saidi_Bondbreak:2007} We
also find this to be true for the {\Cz} molecule, as discussed below.
Although a multi-reference $\ket{\Phi_r}$ containing $N_D$ determinants
increases the computational cost roughly by a factor of $N_D$, we have found that
this is significantly offset by the gain in statistical and systematic
accuracy due to the use of a $\ket{\Phi_r}$ that more closely resembles the
ground state. \cite{Al-Saidi_Bondbreak:2007}

\begin{figure*}
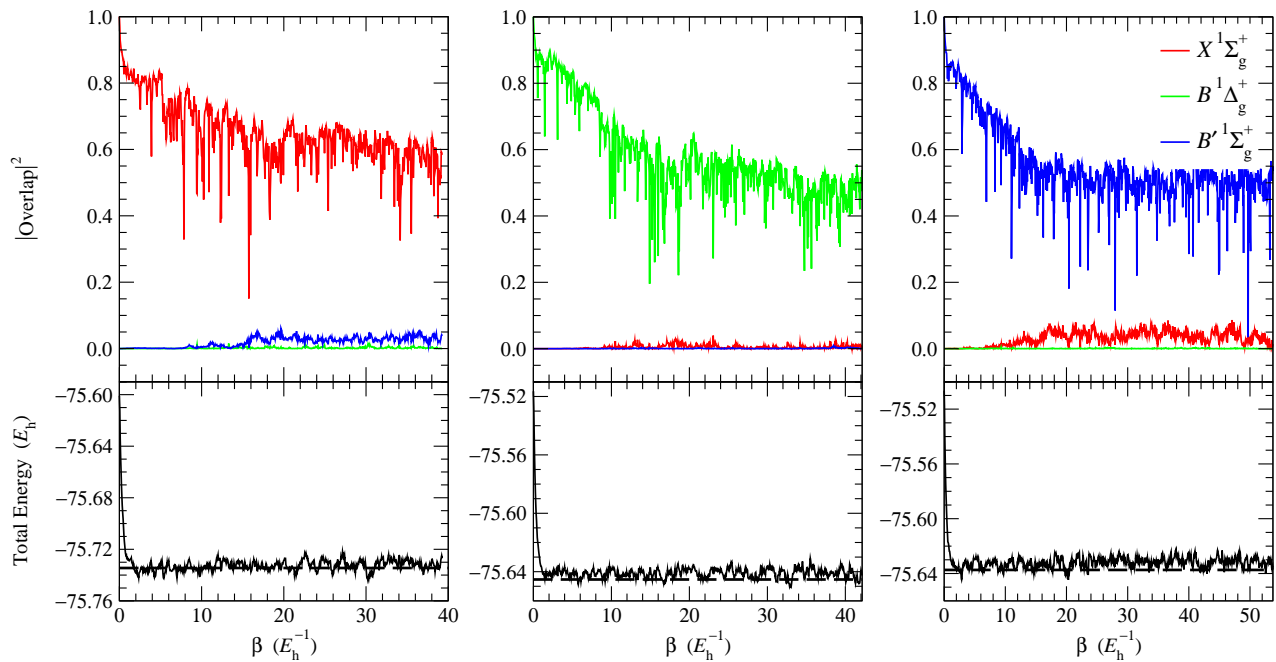


\includegraphics[scale=0.45]{eqlb-ovlp-QMC-CAS816-X.eps}\;%
\includegraphics[scale=0.45]{eqlb-ovlp-QMC-CAS816-B.eps}\;\;%
\includegraphics[scale=0.45]{eqlb-ovlp-QMC-CAS816-Bp.eps}

\caption{
  (Color online)
  The overlaps of the AFQMC wave function with different states, and 
  the computed total energy, as a
  function of imaginary time $\beta$ at interatomic separation $1.25$
  {\AA} in {\Cz}. The left, center, and
  right panels show results for the $X$, $B$, and $B'$ states,
  respectively, with AFQMC using the corresponding truncated CASSCF(8,16)
  wave function as $\ket{\Phi_r}$. In each case, the upper panel shows the
  three overlap integrals
  $|\braket{\Phi_r^{X}}{\Psi_{\mathrm{MC}}}|^2$,
  $|\braket{\Phi_r^{B}}{\Psi_{\mathrm{MC}}}|^2$, and
  $|\braket{\Phi_r^{B'}}{\Psi_{\mathrm{MC}}}|^2$.
The lower panels show the computed AFQMC
  energy, given by Eq.~(\ref{eq:Emixed}), together with the 
exact FCI energy~\cite{Abrams2004} for comparison (indicated by the
  horizontal dashed lines).
 The
  $\ket{\Phi_r}$ are given by truncated CASSCF(8,16) wave functions.
The 6-31G* basis set is used. 
}
\label{fig:overlap}

\end{figure*}

\subsection{Excited states}
\label{sec:excited-states}

Calculating the energy of the lowest excited state belonging to an
irreducible representation that is different from that of the ground
state (\eg{the $B\,^1\Delta_g^+$ state}) is straightforward. In this case, we simply choose
$\ket{\Phi_r}$ in \Eq{eq:Emixed} to have the symmetry of the desired
excited state.
Since the Hamiltonian is invariant under the group of
symmetry transformations, this projects out, in \Eq{eq:El}, any
component in the walker determinant $\ket{\phi}$ belonging to a
different representation. Although it is usually not possible for a
single-reference $\ket{\Phi_r}$
to satisfy the symmetry requirement, multi-reference
wave functions can, at least approximately. We use truncated
complete active space self-consistent field (CASSCF)~\cite{casscf} 
wave functions in this study, and
our tests indicate that symmetry breaking due to the truncation
is small (see below).

For calculations of excited states belonging to the same
irreducible representation as the ground state, \eg{the $B'\,^1\Sigma_g^+$
state}, we rely on the fact that the corresponding reference
wave function is approximately orthogonal to the exact ground state,
$\braket{\Phi_r^{B'}}{\PsiGS^X} \approx 0$.
Obtaining accurate AFQMC results for excited states thus depends on using sufficiently accurate 
excited state trial wave functions. 
Our results indicate that a multi-reference $\ket{\Phi_r}$ with a modest 
number of determinants directly taken from a CASSCF calculation is adequate. 

Figure~\ref{fig:overlap} illustrates our approach for 
the first three singlet states in
{\Cz} at a bond length near that of the ground state equilibrium.
In AFQMC the wave function is given by
\begin{equation}
\ket{\Psi_{\mathrm{MC}}} = \widetilde{e^{-\beta\Hop}} \ket{\Phi_r} 
\sim \sum_\phi w_\phi \frac{\ket{\phi}}{\braket{\Phi_r}{\phi}} \,,
\label{eq:Psi_MC}
\end{equation}
where ${\ket{\Phi_r}}$ is the reference function used in the calculation
for the phaseless constraint and for importance sampling, and the 
sum is over the random walker population, $\{w_\phi,\ket{\phi}\}$, 
at each time slice. From Eq.~(\ref{eq:Psi_MC}), we can 
obtain estimates of the overlap integrals
  $|\braket{\Phi_r^{s}}{\Psi_{\mathrm{MC}}}|^2$ 
(where $s=X$, $B$, or $B'$)
to probe the composition of the AFQMC wave function.
We normalize
$\ket{\Psi_{\mathrm{MC}}}$ by explicitly
evaluating $\sqrt{\langle \Psi_{\mathrm{MC}}|\Psi_{\mathrm{MC}}\rangle}$
at each $\beta$.
Because this involves ``undoing'' the 
importance sampling \cite{Purwanto2004} [division by the factor 
$\braket{\Phi_r}{\phi}$ on the right-hand side of \Eq{eq:Psi_MC}],
there are large statistical fluctuations, as can be seen in the upper 
panels. An average population of 1000 walkers is used in these calculations.

The left panel in Fig.~\ref{fig:overlap}
shows the ground state calculation, and its upper panel shows a
large ($\sim 60\%$) ~ $|\braket{\Phi_r^{X}}{\Psi_{\mathrm{MC}}}|^2$
overlap of the AFQMC walker population with the reference wave
function, as expected. By contrast, the overlap
$|\braket{\Phi_r^{B}}{\Psi_{\mathrm{MC}}}|^2$ is essentially zero,
since the symmetry of the $\ket{\Phi_r^{B}}$ state is different from that
of the $\ket{\Phi_r^{X}}$ ground state, and since symmetry breaking due
to truncation of the full CASSCF wave funtion is evidently weak.
Moreover, the somewhat larger $|\braket{\Phi_r^{B'}}{\Psi_{\mathrm{MC}}}|^2$ overlap 
($\sim 2-5\%$) is not surprising, 
since $\ket{\Phi_r^{B'}}$ has the same
$^1\Sigma_g^+$ symmetry as the ground state. The same trends are
observed in the center and right panels of
Fig.~\ref{fig:overlap}. In the center panel, where the total
energy of the $B$ state is calculated, the overlaps of the walker
population with the different symmetry $X$ and $B'$ states is
extremely small. In each of the three panels, the walker population overlap with the
corresponding trial reference state is large. Finally, it is interesting to
note that the characteristic imaginary time ($\sim 20-40 \,
\Eh^{-1}$) to reach the asymtoptic value of the overlaps is much
larger than the equilibration times ($\sim 3\,\Eh^{-1}$) for the
total energy. (Section~\ref{sec:FCIcompare} presents
detailed comparisons with the FCI results.)

\subsection{Computational details}
\label{sec:comp-details}

\begin{figure}
\includegraphics[scale=0.34]{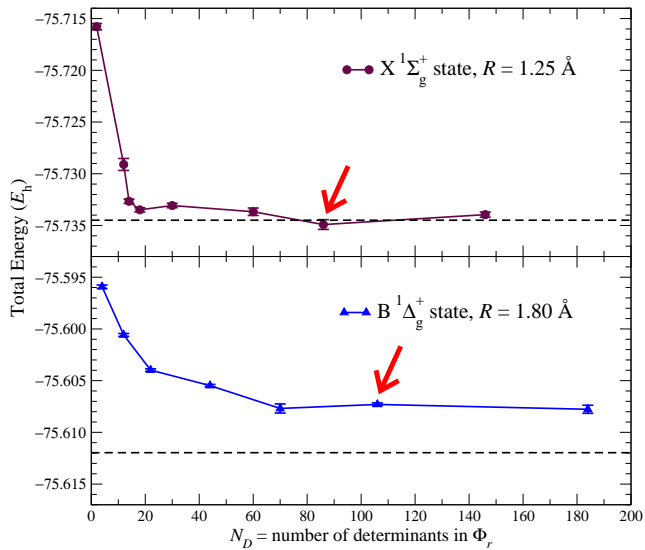}
\caption{
(Color online) Convergence of the AFQMC energy as a function of 
the number of determinants $N_D$ in the multi-reference $\ket{\Phi_r}$,
obtained
by truncating a CASSCF wave function (see text).
The upper panel shows the convergence for the $X$ state near its equilibrium at $R = 1.25$ \AA.
The lower panel corresponds to the $B$ state at $R = 1.8$ {\AA}, a slightly larger bond length
than the $X$-$B$ crossover point at $R \sim 1.7$ \AA).
The arrows indicate the $N_D$ given by our truncation criterion of 97\%
integrated weight.
The 6-31G* basis set was used.
Exact FCI energy~\cite{Abrams2004} is shown as horizontal dashed lines. 
}
\label{fig:convgEvsND}
\end{figure}

Most of our AFQMC calculations use as $\ket{\Phi_r}$
a truncated 
CASSCF(8,16) wave function, 
obtained from the {\GAMESS} quantum chemistry program. \cite{Gamess}
\COMMENTED{
No further
optimization 
of the resulting reference wave function $\ket{\Phi_r}$ is
done.
}
The CASSCF wave function is truncated such that the weight 
(squared coefficient) of the retained
determinants is $\sim 97\%$ of the total.
Figure~\ref{fig:convgEvsND} plots the computed AFQMC energy as a function of the
number $N_D$ of retained determinants (ordered by decreasing weight) for
the $X$ and $B$ states, calculated at $R = 1.25$ and $R = 1.8$ \AA{}, respectively.
Convergence to the $1$ or $2\,\mEh$ level is relatively quick,
with the more correlated $B$ state showing somewhat slower convergence.
Our truncation criterion above gives 
$N_D= 86$ and $N_D = 106$ determinants for the $X$ and $B$ states, 
respectively.
The corresponding AFQMC energies are seen to be well converged with respect 
to the truncation.\cite{truncated_natural-orbitals}

Our AFQMC calculations use the local energy formalism, \cite{SZ-HK:2003,Purwanto2004}
using standard gaussian-type basis sets. \cite{EMSL_BasisSets2007}
Reference wave functions were obtained using {\GAMESS},\cite{Gamess} and
the one- and two-body matrix
elements were obtained using a modified {\NWCHEM} code. \cite{NWChem-4.6}
A mean-field background subtraction is applied to the Hamiltonian prior to
the HS transformation, which improves the
computational efficiency and reduces systematic errors.
\cite{Al-Saidi_GAFQMC:2006,Purwanto2005}
In most of our AFQMC calculations, we use $\Delta\tau = 0.01\,\Eh^{-1}$.
We confirmed that the resulting Trotter error is less than $1\,\mEh$
by calculations at multiple  $\Delta\tau$ values at
selected geometries in both the 6-31G* and 
larger basis sets.

All runs use an average population of about 100 random walkers,
with initial population generated from the RHF wave
function
or "broken symmetry" RHF.\cite{Abrams2004}
(A short phaseless AFQMC projection is first invoked for $\beta\sim 1\,\Eh^{-1}$,
with the RHF wave function as
$\ket{\Phi_r}$ and its copies to form the initial population; the
resulting population, which is purely spin-singlet,\cite{Purwanto2008}
is then fed into the regular calculation with the CASSCF $\ket{\Phi_r}$.)
Typical runs have an equilibration phase of $\beta\sim 10\,\Eh^{-1}$ and then
a growth phase
of  $\beta\sim 10\,\Eh^{-1}$, in which the
trial energy is adjusted via the growth estimator \cite{sz-cpmc}
and set for the rest of the simulation.
A measurement of $\beta \sim 150\,\Eh^{-1}$ is needed
to achieve a statistical accuracy of $1\,\mEh$.
To give an idea of the present
computational cost, such a calculation with the cc-pVTZ basis at a single geometry
(with $N_D \sim 100$ in the trial wave function)
requires approximately 2.5 days on one core of a 2.2GHz Opteron processor.
The current implementation simply imports\cite{Al-Saidi_GAFQMC:2006} one-
and two-body gaussian matrix elements from other quantum chemistry
programs.
The Hamiltonian and overlap integrals are treated as dense matrices.
No special properties of the underlying gaussian basis set are exploited.

Our focus with the phaseless AFQMC has so far
been on establishing the basic
framework in a variety of systems, and testing
its systematic accuracy and robustness.
In this paper our main purpose is to present a proof of concept for excited state
calculations.
Although the present implementation (with gaussian basis sets and
a density decomposition of the 
two-body interaction)
has shown excellent accuracy, there remains considerable flexibility in
the choice of the one-particle basis and the form of the HS transformation.
It is possible that exploiting the flexibility can lead to significant
further improvement in accuracy and computational efficiency.

\section{Results and Discussion}

As bonds are stretched in the dissociation of molecules,
accurate treatment of strong electronic correlations becomes
important, especially in the intermediate regime when bonds first
begin to break.  The {\Cz} molecule is a particularly challenging example,
and this trend is seen clearly below.
The ground state at equilibrium geometry is already nontrivial,
\cite{2007-WF-optimiz} but as the bond is stretched in the ground state and
across all geometries in the excited states, the systematic
errors grow significantly in all calculations.
We first compare our phaseless AFQMC results with benchmark FCI
calculations.\cite{Abrams2004} 
(More detailed discussion about recent theoretical calculations on {\Cz}
molecule can be found in \Ref{Sherrill2005}.)
Calculated PECs and spectroscopic
constants from realistic basis sets are then presented and
compared with experimental results.

\subsection{Comparison with benchmark FCI results}
\label{sec:FCIcompare}

\begin{figure}
\includegraphics[scale=0.65]{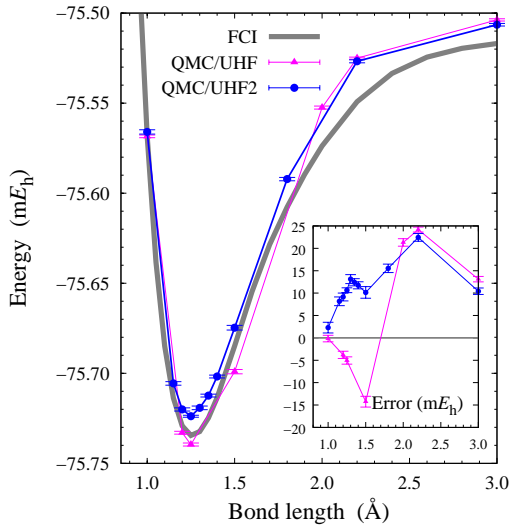}
\caption{
(Color online) {\Cz} ground state PEC: comparison of phaseless AFQMC using
single-determinant UHF and two-determinant UHF2
reference wave functions (see text) with benchmark FCI results from \Ref{Abrams2004}. All calculations used the
6-31G* gaussian basis set.
}
\label{fig:QMC-UHF}
\end{figure}

Figure~\ref{fig:QMC-UHF} compares the phaseless AFQMC ground state PEC
with the FCI results from \Ref{Abrams2004}, using the 6-31G* gaussian
basis set.
In order to benchmark our AFQMC calculations, which do not employ the
frozen core approximation, we estimate a frozen-core correction to FCI
using the difference of UCCSD(T) energies with and without the frozen-core
approximation.
The AFQMC/UHF PEC was calculated using a single
determinant $\ket{\Phi_r}$ from a UHF calculation for the singlet
ground state. The AFQMC/UHF2 PEC used the simplest multi-reference
$\ket{\Phi_r}$ consisting of
two determinants,
\eql{eq:UHF2-def}
{
    \ket{\mathrm{UHF2}}
&   \propto
    \ket{\mathrm{UHF}} +
    \alpha \ket{\mathrm{UHF}_{ax}}
    \,,
}%
\COMMENTED{
    IMPORTANT: the |UHF> and |UHFax> wfns are NOT orthogonal to each
    other, so their variational parameters cannot be sin and cos of an
    angle.
}
where the parameter $\alpha$ is variationally optimized.
In Eq.~(\ref{eq:UHF2-def}), $\ket{\mathrm{UHF}}$ is the usual UHF wave function,
which is a broken symmetry state with opposite spins on the two C atoms, 
but which preserves the axial symmetry of the molecule.
\COMMENTED{
Its valence orbitals can be described as 
$(2s\sigma_g + 2p_z^{(2)})$,
$(2s\sigma_u^* + 2p_z^{(1)})$,
$2p_y^{(1)}$,
$2p_x^{(1)}$,
where the superscript $(1)$ or $(2)$ indicates which atom the
orbital is centered on. 
}
The $\ket{\mathrm{UHF}_{ax}}$ state \cite{note_conf_UHF_ax} 
breaks the axial symmetry of the molecule
and is analogous to the ``antiferromagnetic
solution'' found in a local density approximation calculation in %
Ref.~\onlinecite{Perdew1992}.
\COMMENTED{
to the axially broken RHF wave
function discussed in Ref.~\onlinecite{Abrams2004}.
Its spin-up valence orbitals can be described as 
$2(s-p_z)\sigma_g$,
$[2(s+p_z)\sigma_u^* + 2p_x \pi_g^*]$,
$(2p_z\sigma_g + 2p_x \pi_u)$,
$2p_y\pi_u$.
This configuration is very similar to the the broken ``antiferromagnetic
solution'' found in a local density approximation (LDA) calculation in %
Ref.~\onlinecite{Perdew1992}.
}
Near equilibrium, the energy of $\ket{\mathrm{UHF}}$ is lower than that of
$\ket{\mathrm{UHF}_{ax}}$,
but, as the bond is stretched beyond  $R \sim 1.5$ \AA{}, the $\ket{\mathrm{UHF}_{ax}}$ energy becomes lower.
The combined $\ket{\mathrm{UHF2}}$ state in Eq.~(\ref{eq:UHF2-def}) has a variational energy that
is $\sim 10\,\mEh$ lower than the $\ket{\mathrm{UHF}}$ near equilibrium and
$\sim 20\,\mEh$ lower at $R \sim 1.8$ \AA{}.

The AFQMC/UHF PEC has 
an NPE of $38\,\mEh$ while AFQMC/UHF2 has a smaller NPE of
$20\,\mEh$.
The NPEs of standard quantum chemistry calculations are shown in
Table~\ref{tbl:NPE-qchem}, together with those of AFQMC.
The single reference AFQMC/UHF NPE is thus
seen to be significantly better than RCCSD(T) and comparable to
UCCSD(T).
Using only two determinants, AFQMC/UHF2 has a smaller NPE
than the CISDTQ result.
(AFQMC/UHF2 has a
lower energy at large bond lengths due to the change in the leading
terms of the FCI wave function from the single RHF-like
configuration near equilibrium to a two-determinant configuration in
the dissociation limit.)  Nevertheless, an NPE of $\sim 20\,\mEh$ is
unacceptably large for a high-level method such as QMC. We also note that removing spin-contamination in the walker population,
as discussed in \Ref{Purwanto2008}, does not yield
significant improvements to either AFQMC/UHF or AFQMC/UHF2.
This indicates that these $\ket{\Phi_r}$ are themselves poor.

\begin{table}
\newcommand\fnA {}
\caption{
  Nonparallelity error (NPE) of standard quantum chemistry methods for the
  {\Cz} ground state PEC (taken from \Ref{Abrams2004}), compared with
  that of the phaseless AFQMC method using
  UHF, two-determinant UHF2, and truncated CASSCF(8,16) trial wave functions.
  The range of bond lengths is $R = 0.9$ -- $3.0$ {\AA}.
  All calculations used the 6-31G* gaussian basis set.
}
\label{tbl:NPE-qchem}
\begin{tabular}{lr}
\hline
\hline
Method        & NPE (\mEh) \\ 
\hline
RHF\fnA       & $ 212$ \\
UHF\fnA       & $  78$ \\
MP2\fnA       & $ 130$ \\
RCCSD(T)\fnA  & $  98$ \\
UCCSD(T)\fnA  & $  34$ \\
CISD\fnA      & $ 116$ \\
CISDT\fnA     & $  51$ \\
\vspace{0.2cm}
CISDTQ\fnA    & $  26$ \\
\hline
AFQMC/UHF     & $  38$ \\
AFQMC/UHF2    & $  20$ \\
AFQMC/CASSCF  & $   7$ \\
\hline
\hline
\end{tabular}
\end{table}

\begin{figure}
\includegraphics[scale=0.32]{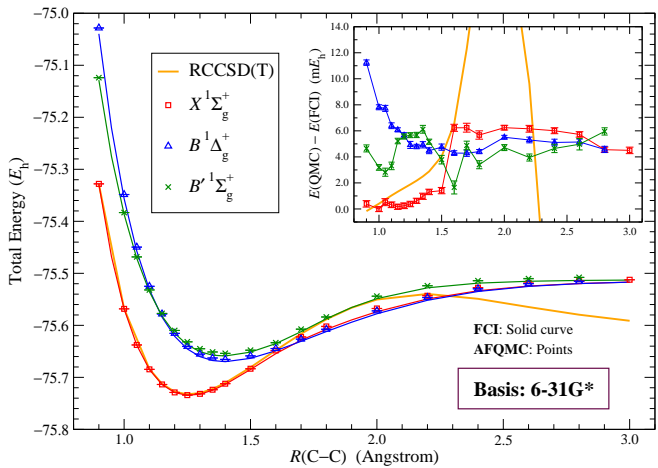}
\caption{
  (Color online) AFQMC/CASSCF PECs for the three lowest lying singlet states
  in {\Cz}, using multi-reference
  truncated CASSCF(8,16) $\ket{\Phi_r}$, compared with FCI
  results.\cite{Abrams2004} FCI results are given by the solid
  curves and AFQMC results are given by symbols with error bars.
  RCCSD(T) results for the ground state are also shown as the orange line.
  The inset shows deviations
  from FCI in {\mEh}. All calculations used the
  6-31G* gaussian basis set.
}
\label{fig:QMC-CAS-6-31Gx}
\end{figure}

We now show that the phaseless AFQMC results become much more accurate
when
multi-reference $\ket{\Phi_r}$ are used.
As described in Sec.~\ref{sec:comp-details},
these are truncated CASSCF(8,16) wave functions,
with no 
further
optimization. 
Figure~\ref{fig:QMC-CAS-6-31Gx}
compares the AFQMC/CASSCF and FCI calculated PECs.
We see that all three PECs are mapped out
accurately, including
the $B'$ PEC which is of the same symmetry as the  ground state $X$.
The overall accuracy of the AFQMC
PECs for both the ground state and the excited states is better than $8\,\mEh$
for all but one point, the smallest bond length in $B$.
The $X$ and $B$ crossover at $R\sim 1.7$~{\AA} and the $B$ and $B'$ crossover
at $R\sim 1.1$~{\AA} are both accurately described.
With the truncated CASSCF trial wave function,
the energies also appear to be variational in all cases, while the
AFQMC/UHF energies are not.
\COMMENTED{We also note from Table~\ref{tbl:NPE-qchem} that the AFQMC/CASSCF NPE
is comparable to the multi-reference perturbation theory CASPT2.
The multi-reference CI (MRCI) method yields much better accuracy;
however it lacks size-consistency and has much steeper scaling than AFQMC.}

Table~\ref{tbl:spectrosc-6-31Gx} presents the spectroscopic constants
corresponding to the PECs in Fig.~\ref{fig:QMC-CAS-6-31Gx}.
Since CCSD(T)
is poor in the dissociation limit, the dissociation energies
$D_e$ in Table~\ref{tbl:spectrosc-6-31Gx} are calculated
using the energy of the free C atom for each method.
For AFQMC, this somewhat improves the comparison of $D_e$ with FCI,
since the AFQMC energy of the C atom is more accurate
than that from the value in Fig.~\ref{fig:QMC-CAS-6-31Gx}
in the dissociation limit, where the truncated CASSCF(8,16)
$\ket{\Phi_r}$ is not very good.
The spectroscopic constants obtained from energy expectation values of the truncated CASSCF $\ket{\Phi_r}$
have substantial fitting uncertainties,
which originate from noise in the determinant truncation.
Both the AFQMC/CASSCF and the full CASSCF results are 
in very good agreement with the exact FCI results, while those based on
energy expectation values of the truncated CASSCF $\ket{\Phi_r}$ are significantly worse.
This  shows that 
AFQMC improves significantly over the truncated CASSCF trial wave functions.

\begin{table}
\caption{
  AFQMC/CASSCF calculated {\Cz} spectroscopic constants, corresponding to Fig.~\ref{fig:QMC-CAS-6-31Gx} using multi-reference
  truncated CASSCF(8,16) $\ket{\Phi_r}$. 
For comparison, results from FCI, \cite{Abrams2004}  CCSD(T), the {\em full\/}
CASSCF, and from the energy expectation values 
of the {\em truncated\/} CASSCF $\ket{\Phi_r}$
are also shown.
  All calculations used the 6-31G* basis set.
  Results for the equilibrium bond length $r_e$ (in \AA),
  vibrational frequency $\omega_e$ (in cm$^{-1}$),
  and ground state dissocation energy $D_e$ (in eV) are shown.
  For excited states, the excitation energy $T_e$ is defined as the
  energy difference between the minima of the excited and ground states.
  Combined statistical and fitting errors in AFQMC are shown in parantheses,
  while pure fitting uncertainties are shown in square brackets.
}
\label{tbl:spectrosc-6-31Gx}
\begin{tabular}{l rrrrr} 
\hline
\hline
 &
\multicolumn{2}{r}{CASSCF(8,16)} \\
 &
{full} &
{truncated} &
{CCSD(T)} &
QMC &
FCI \\
\hline
\multicolumn{2}{l}{$X\,^1\Sigma_g^+$ ground state} \\
\quad$r_e$      &  {1.2563}&  {1.271[4]} &  {1.2577}& {1.2567(3)} & {1.2581}   \\
\quad$\omega_e$ &  {1893}  &  {1705[76]} &  {1869}  & {1888(12)}  & {1863}     \\
\quad$D_e$      &  {6.530} &  {5.12[1]}  &  {5.953} & {6.085(5)}  & {6.030}    \\
\hline
\multicolumn{2}{l}{$B\,^1\Delta_g^+$ excited state} \\
\quad$r_e$      &  {1.3997}&  {1.39[3]}  &          & {1.4004(9)} & {1.4000}   \\
\quad$\omega_e$ & {1414[1]}&  {1360[332]}&          & {1394(13)}  & {1391}     \\
\quad$T_e$      &  {1.712} &  {1.75[8]}  &          & {1.874(6)}  & {1.761}    \\
\hline
\multicolumn{2}{l}{$B'\,^1\Sigma_g^+$ excited state} \\
\quad$r_e$      &  {1.3868}&  {1.399[7]} &          & {1.4009(12)}& {1.3931}   \\
\quad$\omega_e$ &  {1463}  &  {1453[127]}&          & {1325(32)}  & {1394}     \\
\quad$T_e$      &  {1.920} &  {1.91[3]}  &          & {2.189(6)}  & {2.058}    \\
\hline
\hline
\end{tabular}
\end{table}

\begin{figure}[t]
\includegraphics[scale=0.32]{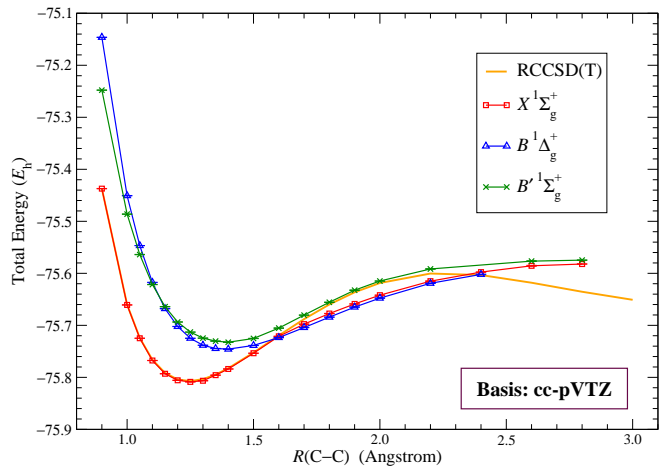}
\caption{
(Color online) {\Cz} AFQMC/CASSCF PECs using $\ket{\Phi_r}$ given
by a truncated CASSCF(8,16) wave function. For comparison, the RCCSD(T) ground
state PEC is also shown.
All calculations used the cc-pVTZ basis. \cite{EMSL_BasisSets2007}
}
\label{fig:QMC-CAS-cc-pVTZ}
\end{figure}

\subsection{Realistic \Cz{} results using large basis sets}

In this subsection, we present phaseless AFQMC/CASSCF PECs with large basis sets and compare
with experimental spectroscopic values.
Figure~\ref{fig:QMC-CAS-cc-pVTZ} shows AFQMC/CASSCF PECs for
the three lowest lying singlet states,
using truncated CASSCF(8,16) wave function as $\ket{\Phi_r}$,
in the cc-pVTZ basis set. \cite{EMSL_BasisSets2007}  For comparison,
the figure also shows the RCCSD(T) ground state PEC, which was
calculated using the {\NWCHEM} \cite{NWChem-4.6} computer program.
Table~\ref{tbl:spectrosc-cc-pVTZ} presents the corresponding
spectroscopic constants and experimental values.
We have also computed the ground state spectroscopic constants
using the larger cc-pVQZ basis set, \cite{EMSL_BasisSets2007}
shown in Table~\ref{tbl:spectrosc-cc-pVQZ}.
As seen from these two tables,
the residual finite basis set error is small and the computed
spectroscopic constants are nearly converged.
(Ground-state results obtained using the
cc-pV5Z basis set by the contracted
multi-reference CI (CMRCI) method \cite{Boggio-Pasqua2000} 
are also included in Table~\ref{tbl:spectrosc-cc-pVQZ}.) 

The calculated AFQMC results in Tables~\ref{tbl:spectrosc-cc-pVTZ} and
\ref{tbl:spectrosc-cc-pVQZ} are in very good
agreement with experimental values.
The CASSCF(8,16) results, {i.e.}, using the \emph{full} CASSCF WF
are also quite
good.
Each CASSCF(8,16) WF consists of $414864$ determinants in the
$D_{2\mathrm{h}}$ space group.
\COMMENTED{This statement is true for any C2 CASSCF(8,16) WF.}
The truncated CASSCF $\ket{\Phi_r}$ used in AFQMC, however, uses only the
first $52 - 275$ determinants, which amounts to $97\%$ of the total
CASSCF determinant weight.
At smaller distances, the correlation effects are small and thus
fewer determinants are needed.
At large bond lengths, 
there are more determinants included
in the $\ket{\Phi_r}$, indicative of stronger correlation effects.
As seen in Table~\ref{tbl:spectrosc-cc-pVTZ}, AFQMC significantly improves
the results when compared to those obtained from the variational estimate
using the \emph{truncated} CASSCF WF.
The level of agreement here between
the AFQMC results with larger basis
sets and experiment is consistent with that in the smaller basis set
between AFQMC and FCI in the previous subsection.

\begin{table}[t]
\caption{
  AFQMC/CASSCF calculated {\Cz} spectroscopic constants compared to experiment. 
  Results from CCSD(T), the {\em full\/}
CASSCF, and from the energy expectation values of
the {\em truncated\/} CASSCF wave functions are also shown.   
  Conventions are as in Table~\ref{tbl:spectrosc-6-31Gx}.
  All calculations used the cc-pVTZ basis set.
}
\label{tbl:spectrosc-cc-pVTZ}
\begin{tabular}{l rrrrrl} 
\hline
\hline
\\
 &
\multicolumn{2}{r}{CASSCF(8,16)} \\
 &
{full} &
{truncated} &
{CCSD(T)} &
QMC &
Expt. \\
\hline
\multicolumn{2}{l}{$X\,^1\Sigma_g^+$ ground state} \\
\quad$r_e$      &  {1.2479}  &  {1.262[3]} &  {1.2508}& {1.2462(9)} & {1.2425}  \\
\quad$\omega_e$ &  {1862}    &  {1766[67]} &  {1842}  & {1884(17)}  & {1855}    \\
\quad$D_e$      &  {6.53}    &   {4.64[1]} &  {6.03}  & {6.32(1)}   & {6.33}    \\
\hline
\multicolumn{2}{l}{$B\,^1\Delta_g^+$ excited state} \\
\quad$r_e$      & {1.397[1]} & {1.416[7]}  &          & {1.391(1)}  & {1.3855}  \\
\quad$\omega_e$ & {1395[23]} & {1351[101]} &          & {1376(23)}  & {1407}    \\
\quad$T_e$      & {1.511[5]} & {1.38[3]}   &          & {1.723(7)}  & {1.498}   \\
\hline
\multicolumn{2}{l}{$B'\,^1\Sigma_g^+$ excited state} \\
\quad$r_e$      &   {1.381}  & {1.398[4]}  &          & {1.393(1)}  & {1.3774}  \\
\quad$\omega_e$ &  {1489[4]} & {1392[63]}  &          & {1441(12)}  & {1424}    \\
\quad$T_e$      & {1.779[2]} & {1.68[2]}   &          & {2.082(8)}  & {1.910}   \\
\hline
\hline
\end{tabular}
\end{table}

\begin{table}[!htb]
\caption{
  AFQMC/CASSCF calculated {\Cz} ground state spectroscopic constants compared to experiment.
  Conventions are as in Table~\ref{tbl:spectrosc-6-31Gx}.
  Calculations used the cc-pVQZ basis set (except CMRCI which used 
  the cc-pV5Z basis).
}
\label{tbl:spectrosc-cc-pVQZ}
\begin{tabular}{l rrrrrrl} 
\hline
\hline
\\
 &
\multicolumn{2}{r}{CASSCF(8,16)} \\
 &
{full} &
{truncated} &
{CCSD(T)} &
{CMRCI}\footnote{Values from analytical fitting in \Ref{Boggio-Pasqua2000}} &
QMC &
Expt. \\
\hline
\multicolumn{3}{l}{$X\,^1\Sigma_g^+$ ground state} \\
\quad$r_e$      &  {1.2452}&{1.262[3]}&  {1.2459}&  {1.2467}& {1.244(1)}  & {1.2425}   \\
\quad$\omega_e$ &  {1868}  &{1759[29]}&  {1852}  &  {1853}  & {1850(21)}  & {1855}     \\
\quad$D_e$      &  {6.57}  & {4.69[1]}&  {6.19}  &  {6.29}  & {6.41(1)}   & {6.33}     \\
\hline
\hline
\end{tabular}
\end{table}

\section{Summary}
\label{sec:summary}

We have shown that molecular excited-state calculations are possible
with the phaseless auxiliary-field QMC method.
Using CASSCF trial WFs, the method
delivers accurate PECs in
the challenging {\Cz}. The computed spectroscopic constants for the
lowest three singlet states, two of which have the same symmetry,
are in very good agreement with experiment.

\begin{acknowledgments}

The work was supported in part by DOE (DE-FG05-08OR23340 and
DE-FG02-07ER46366). H.K.~also acknowlesges support by ONR
(N000140510055 and N000140811235), and W.P.~and S.Z.~by NSF (DMR-0535592). 
Calculations were performed at the Center for Piezoelectrics by
Design, and the College of William \& Mary's SciClone cluster.
We are grateful to Wissam Al-Saidi and Eric Walter for many useful discussions.

\end{acknowledgments}

\bibliography{QMC}

\begin{thebibliography}{40}
\bibitem[{tru()}]{truncated_natural-orbitals}
\bibinfo{note}{The CASSCF natural orbitals were also used to construct an
  alternative determinantal expansion of the full CASSCF wave function, which
  then was truncated in the usual way. For large $N_D \gtrsim 60$, truncating
  the alternative CAS wave function yielded essentially identical AFQMC
  energies as the original truncation. For smaller $N_D$, however, convergence
  of the AFQMC energy was slower than with the original method of truncation.}
\bibitem[{not()}]{note_conf_UHF_ax}
\bibinfo{note}{The spin-up valence orbitals of $\ket{\mathrm{UHF}_{ax}}$ can be
  described as $2(s-p_z)\sigma_g$, $[2(s+p_z)\sigma_u^* + 2p_x \pi_g^*]$,
  $(2p_z\sigma_g + 2p_x \pi_u)$, $2p_y\pi_u$.}
\end{thebibliography}

\ifthenelse{\equal{\PrePrint}{1}}
{
  \printtables
  \newpage
  \listoffigures

  \newpage
  W. Purwanto et al., Fig. 1

  \vskip 1cm
  \noindent
  \includegraphics[scale=0.45]{eqlb-ovlp-QMC-CAS816-X.eps}\;%
  \includegraphics[scale=0.45]{eqlb-ovlp-QMC-CAS816-B.eps}\;\;%
  \includegraphics[scale=0.45]{eqlb-ovlp-QMC-CAS816-Bp.eps}

  \newpage
  W. Purwanto et al., Fig. 2

  \vskip 1cm
  \noindent
  \includegraphics[scale=0.55]{C2-QMC-CASSCF816-detcut.eps}

  \newpage
  W. Purwanto et al., Fig. 3

  \vskip 1cm
  \noindent
  \includegraphics[scale=0.85]{C2-bondbreaking-QMC-test-PsiT2-6-31Gx.eps}

  \newpage
  W. Purwanto et al., Fig. 4

  \vskip 1cm
  \noindent
  \includegraphics[scale=0.55]{C2-QMC-CASSCF816-PES-MAR08.eps}

  \newpage
  W. Purwanto et al., Fig. 5

  \vskip 1cm
  \noindent
  \includegraphics[scale=0.55]{C2-QMC-CASSCF816-PES-MAR08-TZ.eps}

}
{}

\end{document}